**Ali R. Baghirzade**
Master of economics
Junior researcher of the research Institute "Innovative Economics"
Plekhanov Russian University of Economics Moscow, Russia
e-mail: Bagirzade.AR@rea.ru
**Kushbakov B.**
Master of economics Financial University under the Government of the Russian Federation
e-mail: Beko@gmail.com

# ASSESSMENT OF THE EFFECTIVENESS OF STATE PARTICIPATION IN ECONOMIC CLUSTERS

**Annotation.** In the article we made an attempt to reveal the contents and development of the concept of economic clusters, to characterize the specificity of the regional cluster as a project. We have identified features of an estimation of efficiency of state participation in the cluster, where the state is an institution representing the interests of society.

At the moment, the structural problems of the Russian economy are of particular importance, which, we must pay tribute, are recognized by the Russian authorities, who have announced a course of diversification, import substitution and innovative development of the economy. The Ministry of regional development of the Russian Federation suggests moving from a policy of equalizing the economic development of regions to a policy of supporting growth points, and great hopes are pinned on the cluster approach in public policy, which was proclaimed in the Concept of long-term socio-economic development of the Russian Federation until 2020 as one of the key tools for achieving its goals.

The implementation of this approach requires the development of mechanisms for supporting regional cluster projects and methods for evaluating its effectiveness, which is actively underway at the moment, and we are joining these processes with this work. At the same time, high efficiency of state participation in a cluster project is now especially necessary in regions with budget deficits, whose problems are compounded by reduced budget injections.

In the economic literature, the emergence of the concept of "cluster" is usually associated with the name of M. porter, who defines a cluster as "geographically concentrated groups of interconnected companies, specialized service providers, firms in the relevant industries, as well as organizations related to their activities (for example, universities, standardization agencies, and trade associations) in certain areas that compete, but at the same time work together" [1, p. 258].

Undoubtedly, the emergence of the concept of economic clusters and its development were influenced by the works of economists and sociologists who studied the processes of concentration of production since the end of the XIX century, such as A. Marshall, A. Lesh, E. Kropotkin, Heckscher and B. Olin, W.Aizard's research focused on the factors that explain the agglomeration of various sectors of the economy in certain areas, and on the study of the relationship between geographical agglomeration and economies of scale. It should also be noted that the concept of "cluster" was used by Soviet and Russian economic geographers A. P.Gorkin and L. V. Smirnyagin and Swedish businesseconomists K. Fredriksson and L.As early as 1970-s, it was used by Lindmark to denote clusters of enterprises in space [2]. Similar ideas were expressed by Soviet economists in the theory of the formation of territorial production complexes.

Among the main reasons for localization of firms, the following can be distinguished:

1.Geographical proximity significantly reduces transportation and sales costs;

2.The opportunity to benefit from the distribution of costs for maintaining and developing resources shared by several companies;

3.Development of formal and informal communications between companies, which increases the level of trust between counterparties and, as a result, reduces transaction costs (information search, negotiations, control over the fulfillment of obligations, etc.);

4.Close ties, mostly informal, between employees of firms within the same locality contribute to the dissemination of implicit knowledge, that is, those knowledge and experience that cannot be easily formalized and transmitted, and are closely tied to the people who carry them.

5.Development of the local skilled labor market, which reduces the cost of finding and evaluating employees, and improves the" quality " of the labor resource.

6.It is favorable for spreading successful experience in applying new technologies and management methods in the cluster.

The theory of clusters or industrial groups has been continued in the works of M. A. Kropotkin. Enright, who created the theory of a "regional cluster", defining it as an economic agglomeration of firms operating in one or more related sectors of the economy [3]. He believed that competitive advantages are created at the regional level. Several large economic entities that make up the "cluster core" create geographically concentrated demand for monotonous components, labor with appropriate qualifications, services and services of a certain orientation, which are localized around the "cluster core" [4].

However, further research interest in this topic resulted in a rather broad (sometimes very free) interpretation of the concept of "cluster". So, R. Martin and P.Sunli found 10 different definitions of clusters [5]. Naturally, this circumstance makes it difficult to study this concept and allows us to define clusters more as a certain concept, a system of views, than as an economic term.

It should be noted that porter, in his works on competition, proposed not just the concept of new forms of production organization, but also identified clusters as objects of state policy aimed at improving national competitiveness. Porter linked the success of cluster development to the competitiveness of the region and the country as a whole.  It is this practical significance that has made the concept of cluster development of the territory very popular recently, not only in scientific but also in political circles. In the economic development programs of many countries of the world (in Russia, this is the Concept of long-term socio-economic development of the Russian Federation until 2020 [6]) a key role is assigned to the maintenance and development of clusters.

Since the concept of economic clusters itself has been considered by the leadership of various countries as an instrument of practical policy, the cluster is already becoming an object of purposeful creation-both on the part of market participants (putting forward cluster initiatives) and on the part of the state (cluster policy). That is, in other words, the concept of a cluster becomes clearly project based.

Western authors use "cluster initiative" as a term that means an organizational attempt to create or develop a cluster in the region, while domestic authors often use "cluster project", which is even included in regulatory documents of government bodies, for example, in the cluster policy strategy of the Yaroslavl region [7].

A cluster initiative has the characteristic features of a project: uniqueness, limited end goal or range of tasks, limited resources, including time limits. Once the goal is achieved, the project is completed, but the cluster itself continues to exist as a form of cooperation.

We define a cluster project as a long-term target program for the development of a regional economic cluster, which has a mixed or combined content. A cluster project is a coordinated, purposeful set of activities within the framework of joint activities of cluster participants. At the same time, the cluster project should be considered as a mechanism for creative interaction between business and the state, and on the part of the latter - as a platform for implementing economic policy for the development of regions.

Due to the involvement of a large number of participants, the cluster project has a significant impact on the national economy and the population. The state accumulates part of the public wealth, redistributes it to solve socially significant problems, and with the support of a particular cluster project, it should check the validity of allocating resources for its implementation from the point of view of society. The need for such an assessment is justified by the macroeconomic concept of limited resources. Therefore, for the state, evaluating the effectiveness of a project means checking whether it is reasonable from the point of view of society to allocate resources for the implementation of this particular project, given the availability of alternatives.

It should be noted that the presence of several participants in a cluster project causes a discrepancy in their interests, different attitudes to the priority of various activities and subprojects. As a result of the implementation of an investment project, specific cash flows are generated for each participant both at the entrance (in the form of investments) and at the exit.

In a narrow sense, the effectiveness of participation in an investment project of funds from state budgets of different levels is characterized by budget efficiency. However, the state is an institution that represents the public interest, so there is a need to assess not only budget efficiency, but also the socio-economic consequences of project implementation.

In cluster projects, the state seeks to fill in the "market gaps", which means that it is forced to take on social and infrastructure projects, the consequences of which cannot always be taken into account with the help of economic effect (which is often negative). In this case, the assessment of the project's effectiveness comes first, that is, the degree to which the needs of stakeholders are met through the achievement of certain goals.

The evaluation of the project in terms of the interests and benefits of society is expressed through the so-called public or social efficiency. Its peculiarity is that it reflects the non-economic consequences of the project: social and environmental effects that are not always quantifiable.

The task of assessing the social impact is to ensure that, first of all, the project meets social norms, conditions and human rights standards and, at a minimum, does not worsen the situation of any member of society. This condition echoes the Pareto optimum, which means that if the welfare of even one member of society has increased and the welfare of all others has not worsened, then the General welfare also increases. And in the second place, this assessment determines the degree of public utility of the project.

In our opinion, for cluster projects as socially significant, the first step is to assess their socio-economic effects. And in case of unsatisfactory public performance, projects are not recommended for implementation. If a cluster project turns out to be socially useful, then the next step is to evaluate the project through the prism of public finances - budget efficiency.

It should be noted that the achieved social effect can affect the indicators of budget efficiency of the project due to their close relationship. For example, reducing poverty provides savings on social transfers, and improving the health of the population provides savings on temporary disability benefits and disability pensions - but an accurate calculation is not possible here.

We propose to Supplement the existing methods of assessing public effectiveness by determining the project's contribution to achieving the strategic development goals of the region, by comparing the project results with the targets and indicators of the region's socio-economic development strategy. This step will allow us to assess the public feasibility of implementing a cluster project in a broad sense, as the degree to which the desired prospective state of society is achieved.

Summing up, we note that the state, when financing a cluster project, is in a very vulnerable position, depending on the firms participating in the cluster, on their commercial success and their tax integrity. The state withdraws part of the cluster's revenue, but does not participate in its creation, only providing support.

**Acknowledgment**

This article was prepared as part of the government contract as requested by the Ministry of Science and Higher Education of the Russian Federation on the subject formulated as «Structural changes in economy and society as a result of achieving the target indicators of National projects, which provide opportunities to organize new areas of social and economic activity, including commercial, both in Russia and abroad» (project No. FSSW-2020-0010)